\documentclass[final]{elsarticle}

\usepackage{amssymb}
\pdfoutput=1
\newcounter{bla}

\journal{Computer Physics Communications}

\begin{document}

\begin{frontmatter}

  \title{C2x: a tool for visualisation and input preparation for {\sc
      Castep} and other electronic structure codes}

\author[a]{MJ Rutter\corref{author}}
\cortext[author] {Corresponding author.\\\textit{E-mail address:} mjr19@cam.ac.uk}

\address[a]{TCM, Cavendish Laboratory, University of Cambridge, JJ Thomson Av,
Cambridge, CB3~0HE. UK}

\begin{abstract}
  The c2x code fills two distinct roles. Its first role is in acting
  as a converter between the binary format {\tt .check} files from the
  widely-used {\sc Castep}[1] electronic structure code and various
  visualisation programs. Its second role is to manipulate and analyse
  the input and output files from a variety of electronic structure
  codes, including {\sc Castep}, {\sc Onetep} and {\sc Vasp}, as well
  as the widely-used `Gaussian cube' file format. Analysis includes
  symmetry analysis, and manipulation arbitrary cell transformations.
  It continues to be under development, with growing functionality,
  and is written in a form which would make it easy to extend it to
  working directly with files from other electronic structure codes.

  Data which c2x is capable of extracting from {\sc Castep}'s binary
  checkpoint files include charge densities, spin densities,
  wavefunctions, relaxed atomic positions, forces, the Fermi level,
  the total energy, and symmetry operations. It can recreate {\tt
    .cell} input files from checkpoint files. Volumetric data can be
  output in formats usable by many common visualisation programs, and
  c2x will itself calculate integrals, expand data into supercells,
  and interpolate data via combinations of Fourier and trilinear
  interpolation. It can extract data along arbitrary lines (such as
  lines between atoms) as 1D output.

  C2x is able to convert between several common formats for describing
  molecules and crystals, including the {\tt .cell} format of {\sc
    Castep}. It can construct supercells, reduce cells to their
  primitive form, and add specified k-point meshes. It uses
  the spglib library[2] to report symmetry information, which it can
  add to {\tt .cell} files.

C2x is a command-line utility, so is readily included in scripts. It
is available under the GPL and can be obtained from {\tt
  http://www.c2x.org.uk}. It is believed to be the only open-source
code which can read {\sc Castep}'s {\tt .check} files, so it will have
utility in other projects.
\end{abstract}


\end{frontmatter}

{\bf PROGRAM SUMMARY}

\begin{small}
\noindent
{\em Program Title: c2x}                \\
{\em Licensing provisions: GPLv3 }      \\
{\em Programming language:} C           \\
{\em Lines of code:} c. 10\,000  \\
{\em Required libraries:} None required. Spglib optional but recommended \\
{\em Platform:} any supporting ANSI C \\
{\em Expected run time:} seconds \\
{\em Primary site for distribution and documentation:} {\tt
  www.c2x.org.uk} \\

{\em Nature of problem:}\\
C2x is able to extract a large variety of data from {\sc Castep}'s[1]
large, binary--format, output files to facilitate further processing
or visualisation. These binary files are optimised for rapid input and
output by {\sc Castep}, but are impossible to read without detailed
knowledge of their internal structure, which varies between different
versions of {\sc Castep}. In them wavefunctions are stored as plane
wave coefficients, so need to be Fourier transformed before they can
be visualised in real space. C2x can manipulate the input files of
{\sc Castep}, and other common crystallographic formats, performing
functions useful for setting up calculations, such as supercell
construction, format conversions, and symmetry analysis.

Different electronic structure codes use different file formats for
outputting densities. C2x is able to convert between some of the major
formats, allowing visualisation and post-processing tools targetted at
one electronic structure code to be used with others.\\
{\em Solution method:}\\
C2x is a command-line utility written in standard C. It is able to
process large (multi-GB) binary files efficiently,
extracting much smaller datasets. C2x has considerable knowledge of
the structure of the checkpoint files as written by various versions of
{\sc Castep}. Wavefunctions are optionally converted to densities, and
then transformed to real space using its own FFT routine. Weighted
sums of densities from multiple bands can be accumulated.
C2x interpolates volumetric data by using user-specified
combinations of trilinear and Fourier interpolation. For symmetry
analysis it relies on the existing spglib[2] library, which is also
used by {\sc Castep}. It supports cell transformations with the
axes of the new cell expressed in absolute terms, or in terms of the
original axes. It has its own internal representation of the unit cell
and its contents, and has several routines for converting this to and
from common crystallographic file formats.

C2x is able to read densities from {\sc Castep}'s binary formats, and
both read and write {\sc Castep}'s formatted density files, {\sc
  Vasp}'s formatted density files, and the {\tt .cube} files used by
Gaussian and {\sc Onetep}.\\
   \\

\end{small}

\section{Introduction}
\label{}

Programs such as {\sc CASTEP}\cite{CASTEP} calculate the electronic
properties of systems, but, in order to gain insights from their
results, it is necessary to have a means of presenting or
post-processing their output. It is also useful to have a mechanism
for assisting with the creation of their input files.

The output {\tt .check} file from {\sc Castep}, which contains the
charge density, and the wavefunctions of all bands at all k-points,
amongst other data, is written in a binary format which preserves full
precision whilst remaining as compact, and as rapid to read, as
reasonably possible. The resulting file can be many gigabytes, and is
unreadable without detailed knowledge of its structure, the detail
of which changes between different {\sc Castep} versions. The c2x
code is able to extract data from this file into formats compatible
with many visualisation programs, and does so without reading
the whole file into memory, so works well on modest computers. It fits
in with a workflow of performing a calculation on a remote
supercomputer, extracting data on a modest login node, and then
transferring the small dataset extracted to a local computer for
visualisation. Alternatively one can transfer the whole output to the
local computer, and then run c2x.

Visualisation is not the only form of post-processing, and c2x's
ability to convert between the output formats of {\sc Castep}, {\sc
  Onetep}\cite{onetep} and {\sc Vasp}\cite{VASP} makes it easy to use
post-processing tools written for one of these programs with output
from the others.

The commercial Material Studio\cite{MatStud} is able to read {\tt
  .check} files directly, but is restricted to the Microsoft Windows
platform, whereas c2x is platform agnostic. When c2x was first created
over a decade ago, then called {\tt check2xsf}, it simply converted
from {\sc Castep}'s {\tt .check} format to XCrysden's XSF
format\cite{xcrysden}, XCrysden being a free, multi-platform, crystal
structure viewer with the ability to view volumetric data, such as
charge densities, along with atoms and bonds. By the time the first
version of {\tt check2xsf} was publicly released in 2007 it had
gained the ability to perform many transformations on {\sc Castep}'s
{\tt .cell} input files.

Functionality has been extended considerably since, an early extension
being the ability to extract spin densities, this being used in
studies of Fe(II)\cite{BJR2009}. Now it provides for a range of
processing of the input and output of {\sc Castep} (and similar
codes). It does not perform any visualisation itself, but its output
is compatible with XCrysden, VMD\cite{vmd}, Jmol\cite{jmol},
{\sc Vesta}\cite{VESTA}, Gnuplot\cite{gnuplot} and similar codes, for it
supports the Gaussian {\tt .cube} format\cite{gaussian}, as well as
XCrysden's {\tt .xsf} format.  It is a command-line utility, so is
easy to use from scripts. This has led to it being extensively used in
both {\it ab initio} random structure searching (AIRSS)\cite{AIRSS},
and also in the {\sc Caesar} electron-phonon coupling
code\cite{caesar}.

C2x converts between various molecule and crystal description file
formats in a similar fashion to Open Babel\cite{babel}. It can read
and write {\sc Castep}'s {\tt .cell} format, including those
extensions introduced by {\sc Onetep}, the {\sc chg}, {\sc chgcar} and
{\sc poscar}
formats of VASP~5, PDB\cite{PDB}, Shelx-97\cite{shelx97},
and subsets of both XSF and CIF\cite{CIF}. It can also write
CML\cite{CML}, a python dictionary, and a format compatible with
python's ASE module\cite{ASE}. Full support for the CIF file format is
unnecessary as the ability to convert from the CIF to the {\tt .cell}
format is already provided by the CIF2Cell code\cite{cif2cell}.

C2x regards the {\tt .cell} format as being four distinct
formats, depending on whether atomic positions are given in relative
or absolute co-ordinates, and whether the cell axes are expressed in
Cartesian form or as $a,b,c,\alpha,\beta,\gamma$. It can convert
between these. {\sc Onetep}'s input file format is very
similar to the {\tt .cell} format, and is supported as a subformat.

Unlike Open Babel, c2x can construct supercells, which has many uses
such as prior to manually removing or replacing an atom to form a
defect, or for use in phonon calculations. Densities can be expanded
to supercells. It can also add k-points using regular grids with
offsets, such as Monkhorst-Pack grids\cite{MP}. Provided that it was
built with spglib\cite{spglib}, it can perform symmetry analysis. This
is the same symmetry analyser which {\sc Castep} uses from version
8.0. In the absence of spglib, it has a simplistic, yet often
effective, algorithm for finding primitive cells.

The functionality of c2x extends beyond simple data extraction
and conversion. It can not only convert reciprocal space wavefunctions
to real space, but it can interpolate data via both tri-linear and
Fourier interpolation, extend volumetric data to supercells (or
transform to other cells), and it can calculate failure stars\cite{MP}
of k-point sets.

Documentation is available both in the form of a standard UNIX man
page, and also from the website {\tt http://www.c2x.org.uk}.

\section{Visualisation of 3D grid data}

\begin{figure}
\begin{center}
\includegraphics[width=0.45\textwidth]{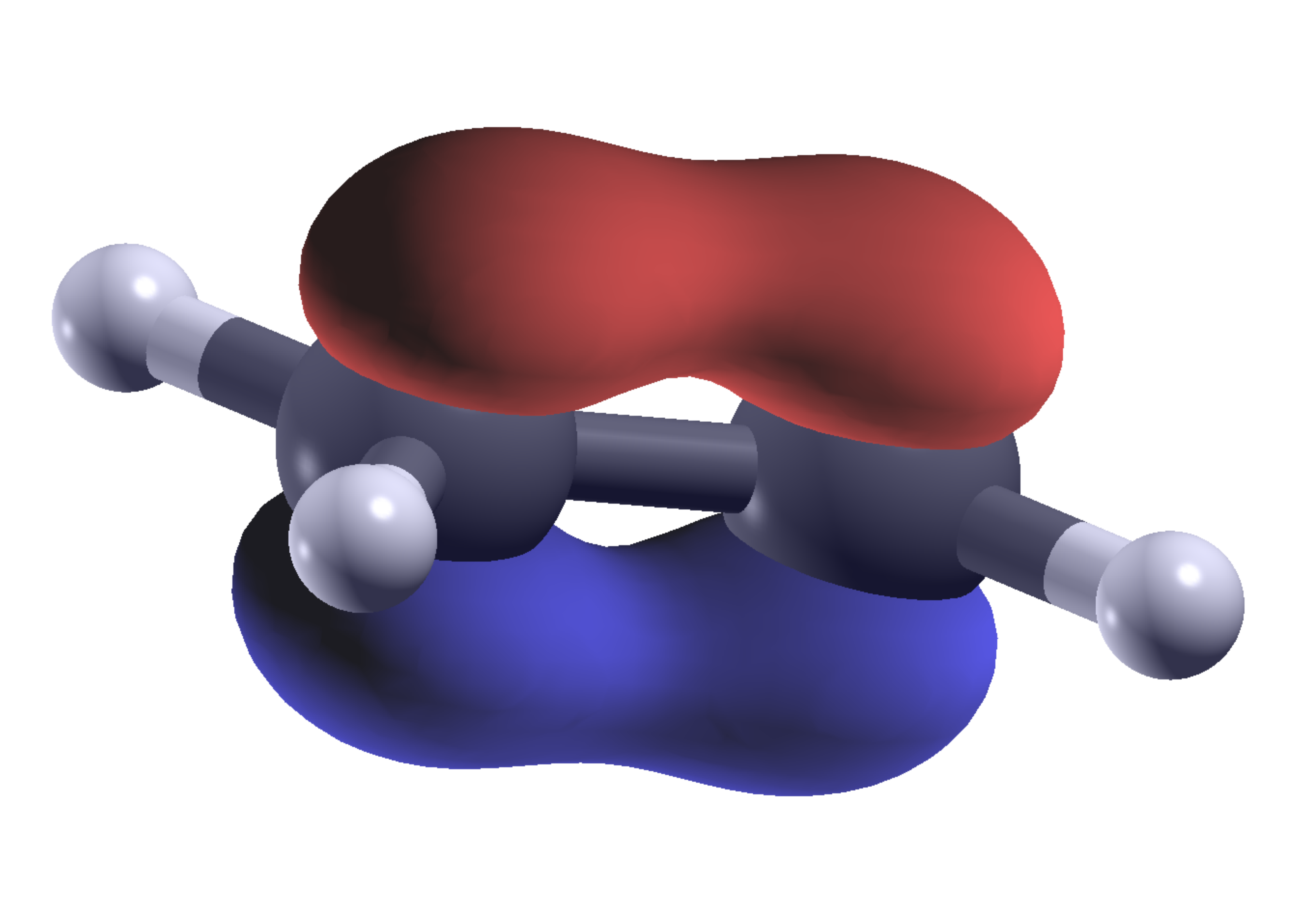}
\hspace{0.03\textwidth}
\includegraphics[width=0.5\textwidth]{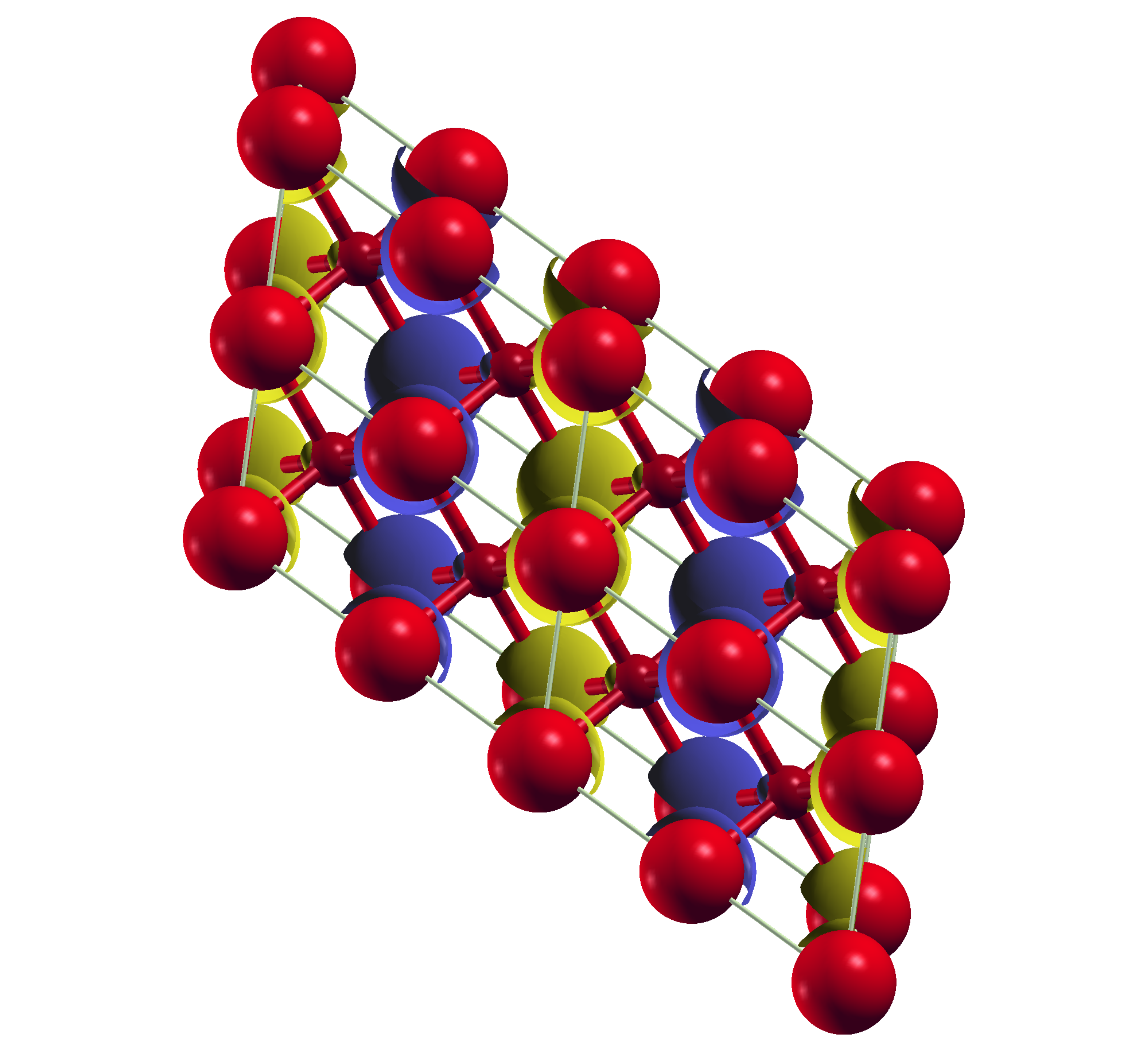}
\end{center}
\caption{Isosurfaces of the $\pi$-bond wavefunction in ethene, and the
  antiferromagnetic spin density in FeO, plotted using XCrysDen after
  extracting the data with c2x. For the FeO spin density, the blue and
  yellow shells around alternate layers of large, red, Fe atoms show
  the alternating spin density.\label{fig:3D}}
\end{figure}

C2x can extract from {\sc Castep}'s {\tt .check} files charge and spin
densities, and wavefunctions. For {\sc Castep} it can read {\tt .castep\_bin} files,
which are similar to {\tt .check} files but lack the wavefunctions,
{\tt .cst\_esp} electrostatic potential files, {\tt .chdiff}
charge density difference files, and {\tt den\_fmt} formatted density
files. It can also read charge and spin from
VASP~5's {\sc chgcar} format. Additionally it
can read a single 3D dataset from a file in the `Gaussian Cube'
format, a volumetric output format now used by electronic structure
programs other than Gaussian\cite{gaussian}, including {\sc Castep}
and {\sc Onetep}. The cube format can be written by c2x, and is a
convenient input format used by VMD and Jmol amongst other programs.

C2x is designed to produce output compatible with XCrysden, Jmol, VMD
and {\sc Vesta}. For 1D data it produces output in a general
two-column format compatible with gnuplot and many other plotting
programs. The ends of the line along which 1D data are required can be
described either in fractional co-ordinates or in terms of atomic
positions.  To produce 1D data, interpolation is almost always
required (unless the points of the line precisely co-incide with
points on the real-space grid). C2x will perform trilinear
interpolation, after first optionally performing Fourier
interpolation. The ability of c2x to produce charge densities and
densities from individual bands along a bond was used by Monserrat and
Needs in their study of electron--phonon coupling in diamond structure
materials\cite{BM2014}.

C2x will also output scalar data, such as the integral of a band or
density, or its global extreme values, or a value at a given
point. For the spin density visualisation of FeO shown in
figure~\ref{fig:3D}, it reports that the maximum value of the spin
density was $8.45$, the minimum $-8.22$, and the sum $-0.05$ (over
25,920 grid points). Given that the FFT grid may be breaking the
symmetry of the system, and the system may not be fully converged,
this is consistent with a pure antiferromagnetic phase with zero net
spin. C2x will also calculate these quantities after Fourier
interpolation (see below). Fourier interpolation will preserve the
integral of the data, but may change the values of the extrema. After
Fourier interpolation onto a grid of twice the original density, c2x
reports extrema of $8.45$ and $-8.45$.

\section{Interpolation}

In common with other plane-wave codes, {\sc Castep} makes several
approximations in order to reduce the complexity of a calculation. Two
notable ones are the use of a finite cut-off for the plane-wave basis
set, and the use of a finite FFT grid. These two approximations are
linked. The FFT grid could be chosen to be just sufficient to store
all Fourier components of the charge density arising from the basis
function, which, being constructed as $|\psi(\mathbf{r})|^2$, will
have components extending to twice the frequency of the basis set.
(The potential is a non-polynomial function of the density, so
must be assumed to have components at all Fourier frequencies.)

\begin{figure}
\begin{center}
\includegraphics[width=0.55\textwidth]{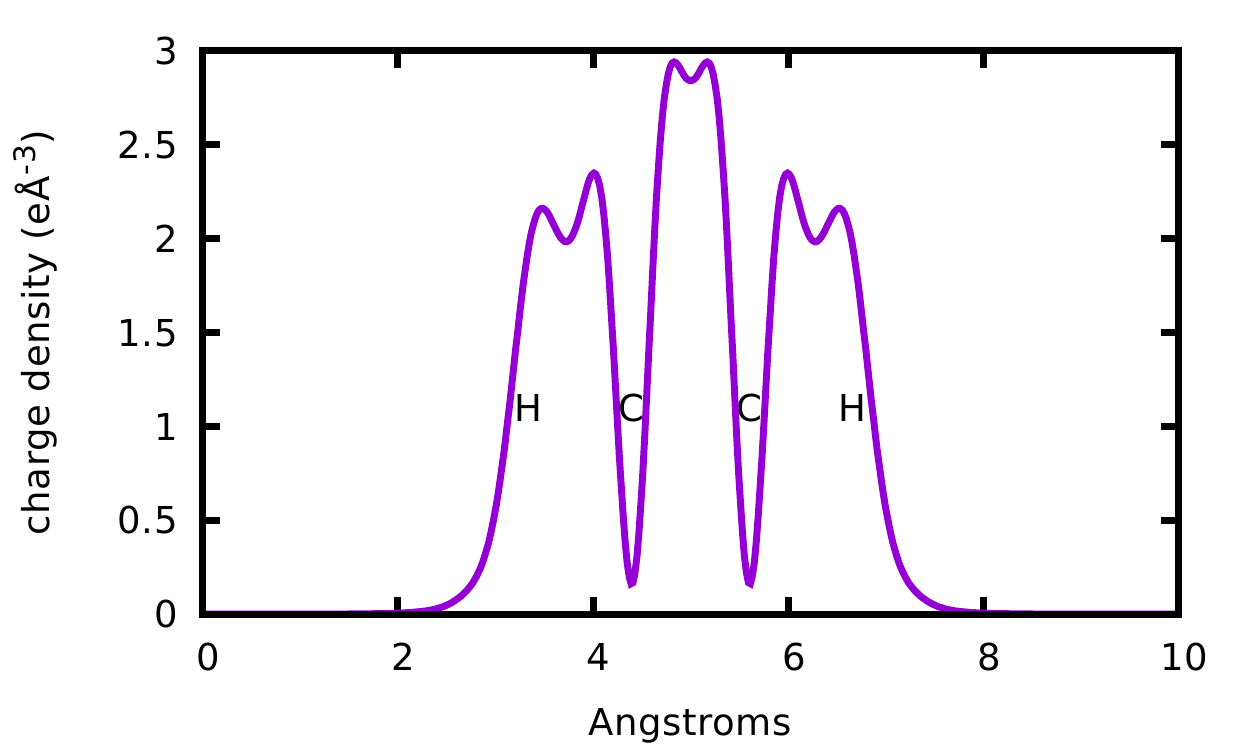}
\end{center}
\caption{Valence charge density along the axis of an acetylene
  molecule in a box.
  The positions of the atomic nuclei are shown. A fine FFT grid was
  used for this calculation.
\label{fig:interp}
}
\end{figure}

\begin{figure}
\begin{center}
\includegraphics[width=0.45\textwidth]{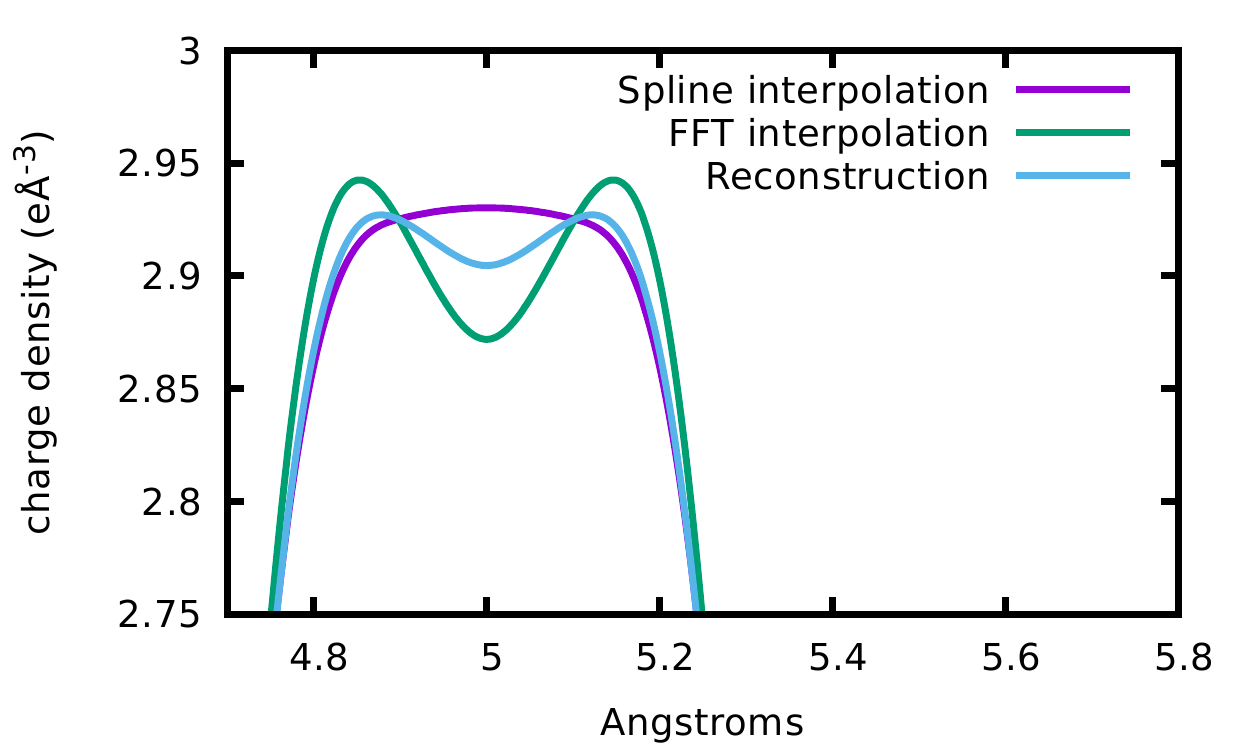}
\hspace{0.05\textwidth}
\includegraphics[width=0.45\textwidth]{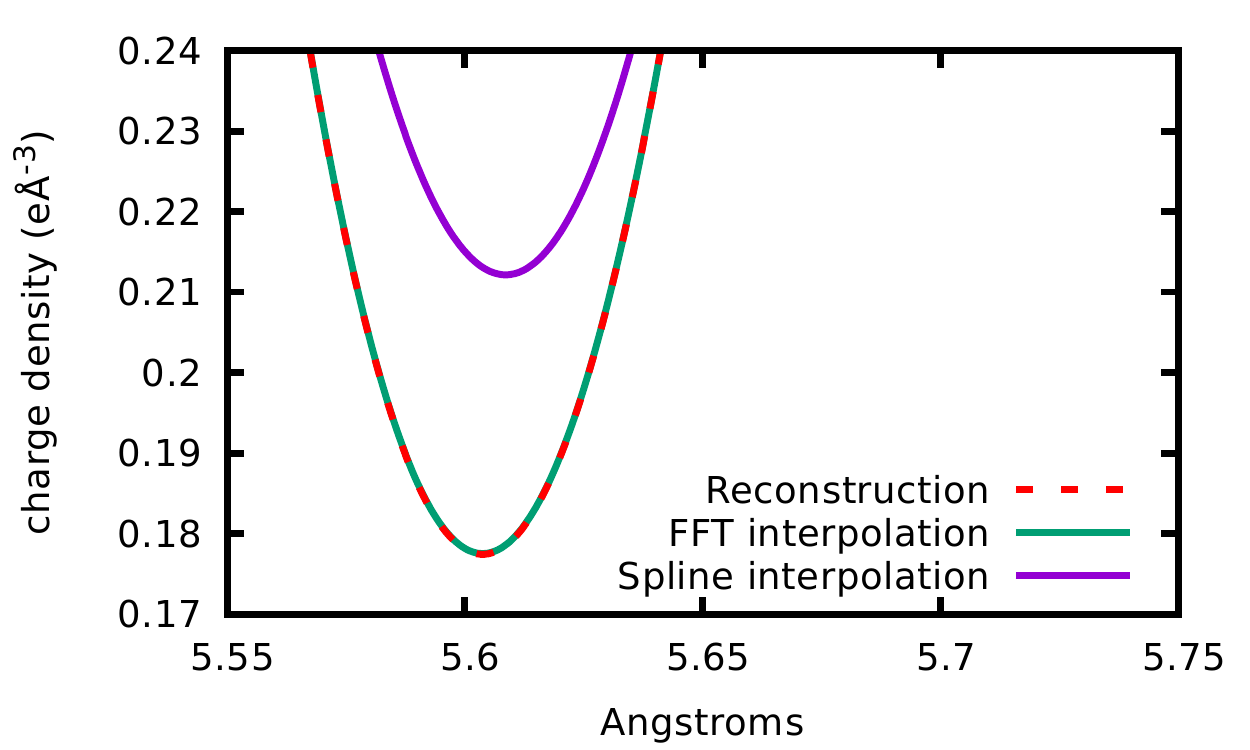}
\end{center}
\caption{Two details of interpolation of the valence charge density in
  acetylene. In the first, focussed on the centre of the C--C bond, a
  very coarse grid shows a qualitative difference between the two
  interpolation methods -- spline interpolation lacks the central
  minimum. In the second, focussed on the orthogonalisation minimum 
  of a carbon atom, a less coarse grid shows that
  Fourier interpolation (bottom curve) is now orders of magnitude more
  accurate than spline interpolation (top curve). At this scale, the
  Fourier-interpolated and reconstructed data are indistinguishable.
\label{fig:interp2}
}
\end{figure}

{\sc Castep} defaults to using an FFT grid which can contain all
Fourier components up to seven-eighths of the largest in the
density, an approximation which is generally considered adequate. It
reduces the number of points in the 3D grid by about a third, with a
corresponding increase in the speed of the FFTs using this grid, and
of operations such as application of potentials. It is quite possible
to perform calculations with a coarser grid than this.

As an example for investigating interpolation schemes, the charge
density along the axis of an acetylene molecule is
considered. Norm-conserving pseudo-potentials are used, and the
calculation is not symmetrised, so that the valence charge density is simply
the sum of the squares of the moduli of the converged wavefunctions of
the individual bands. Figure~\ref{fig:interp} results from plotting
the valence charge density along the axis, interpolated with cubic
splines by Gnuplot, after using a fine, 650eV, cut-off for the plane
waves, and {\sc Castep}'s default FFT grid. The unit cell in which the
molecule is placed is a box 10\AA{} long in the direction of the
molecule's axis. As expected, strong minima are seen where the
wavefunction must be orthogonal to the carbon 1s core electrons, and
slight minima are seen in the centres of the three co-valent bonds.

The left-hand part of figure~\ref{fig:interp2} shows attempts to
interpolate from a very coarse FFT grid. The cut-off has been
reduced to 490eV, and the number of grid points
in the direction of the molecule's axis is just 49. The figure shows
the detail of the C--C bond. Compared to the exact result given by
reconstructing the charge density from the stored wavefunction
co-efficients, cubic spline interpolation is seen to be quantitatively
more accurate, but FFT interpolation (onto a grid of 98 points in the
direction of the molecule's axis) followed by cubic spline
interpolation, is qualitatively better, in that it reproduces the
minimum which is otherwise missed. It is not argued that this
qualitative difference in favour of Fourier interpolation shows that
Fourier interpolation is superior in this case, given the quantitative
difference in favour of spline interpolation.

The method of Fourier interpolation Fourier transforms the real-space data
to reciprocal space, pads onto a larger reciprocal space grid by
setting the new high-frequency components to zero, and then inverse
Fourier transforms back to the corresponding, finer, real-space grid
after appropriate scaling.

Fourier interpolation will be exact if the function interpolated
contains no frequency components above the Nyquist frequency of the
original grid. Its errors will be related to the weight of the
components above the Nyquist frequency. {\sc Castep} calculations are
usually performed in a regime where there is very little weight in the
frequency components of the charge density above the Nyquist frequency
of the FFT grid. If this were not so, the aliasing of the high
frequency components would lead to a loss of accuracy within the
calculation itself. Under this condition, Fourier interpolation might
be expected to be better than cubic spline interpolation, because it
is able to make extra, correct, assumptions about the data (periodic,
and lacking high-frequency components).

The right-hand part of fig~\ref{fig:interp2} shows a detail from the
orthogonalisation minimum at one of the carbon atoms, and, again, two
different interpolation schemes, cubic interpolation, and Fourier
interpolation onto a double density grid followed by cubic
interpolation. The calculation is now being performed with a cut-off
of 550eV and with 60 points in the FFT grid in the direction parallel to
the molecule's axis. The difference in the depth of the minimum
between the cubic interpolation and the reconstructed data is
apparent, and is about 0.035e\AA{}$^{-3}$. The position of the minima
also differ very slightly, by about 0.004\AA{}. On this scale, the
difference in value at the minimum between the Fourier
interpolated and the reconstructed data is not visible. It is a little
under 0.0001e\AA{}$^{-3}$, so about 350 times as accurate as the spline
interpolation. A similar result is obtained by looking at the (two)
global maxima, where spline interpolation gives a value
0.0038e\AA{}$^{-3}$ too large, whereas the difference between Fourier
interpolation and the reconstructed data is under 0.00002e\AA{}$^{-3}$,
so again Fourier interpolation is more accurate by a factor of a
couple of hundred.

In this case Fourier interpolation was performed to interpolate from a
60 point grid to a 120 point grid in the direction along the bond. The
acetylene molecule was placed along the {\bf a} axis, so the c2x
command line for generating the interpolated plot was simply

\begin{verbatim}
c2x -c -i=120,0,0 -P='(0,0,0):(1,0,0):121' ac.check ac_i.dat
\end{verbatim}

meaning extract charge density ({\tt -c}), interpolate onto a grid of
120 points along the {\bf a} axis, without changing the grid density
in {\bf b} or {\bf c} ({\tt -i=}), and output data along a line of 121
points from the origin to (1,0,0) ({\tt -P=}). As the 121 points
include both end points, this is precisely equivalent to a 120 point
FFT grid.

To recalculate the charge density from the stored wavefunctions with
c2x, one simply replaces the {\tt -c} above was replaced by {\tt -BAW}
meaning extract bands as densities ({\tt -B}), accumulate them rather
than extracting separately ({\tt -A}), and weight by occupancy and
k-point weight whilst accumulating ({\tt -W}).

This demonstrates that Fourier and cubic spline interpolation can give
results which are quantitatively different, and that both theory, and
the example given, suggest that Fourier interpolation should be
preferred for reasonable grid resolutions. This ability of c2x to use
different interpolation methods to see what differences result is
valuable.

\section{Cell manipulation}

When performing calculations on defects, surfaces, or phonons, it is
often useful to generate cells of an unusual shape, or cells
which contain multiple repetitions of the original cell provided. This
c2x readily does: one simply provides a cell, and the new axes
either in absolute or fractional co-ordinates. Impossible
transformations are detected and result in an error.

C2x can also regenerate a {\tt .cell} file, including relaxed atomic
positions, from a {\tt .check} file.

When compiled with spglib, c2x exposes various features of that
library, including symmetry determination, primitive cell
determination, and snapping atoms to their precise symmetry
positions. If initial spin states are given in a {\tt .cell} file,
then atoms of the same species but with different spins are considered
to be distinct, and atoms with different {\sc Onetep} labels are also
considered to be distinct. C2x also produces a human-readable
description of a symmetry operation from a symmetry matrix, in terms
of a $n$-fold rotation about an axis plus a translation.

C2x can determine the symmetry of a structure relaxed by {\sc
  Castep} as well as a structure to be used as input.

\section{Structure of Code}

C2x is designed to be easy to compile, and easy to maintain. It is generally
straight-forward to add support for an additional file format by
writing a single routine to read to, or output from, its internal
data structures. A recent format added was the ability to produce output
compatible with python's ASE module. This required a new routine of
under 40 lines, and just six lines added to the main program to parse
the new option, to call the new routine, and to add a line to the help
text. One extra line was also needed in the global header file.  The
routines which read {\sc Castep}'s binary-format files have been
designed to read files of either endianness: host or not host.

The code stores data internally in a set of C structs. Different
structs store different data, such as the unit cell (storing cell
vectors), the contents (storing atoms, an optional title, and
optionally forces), and a linked list of grids, which store 3D grid
data, each with its own grid resolution and title. Routines for
reading and writing take as arguments as many of these as are relevant
to their file format. The {\sc Castep} {\tt .check} file contains all data
relevant to a calculation, so its reader has the most complicated
prototype of

\begin{verbatim}
void check_read(FILE* infile, struct unit_cell *c,
        struct contents *m, struct kpts *kp, struct symmetry *s,
        struct grid *gptr, struct es *elect, int *i_grid);
\end{verbatim}

Here {\tt infile} is a pointer to the file to be read, the {\tt es}
structure specifies which bands (if any) to read, and if Fourier
interpolation is requested {\tt i\_grid} will store the requested
grid. The other structs are output parameters.

In contrast, the reader for pdb files, which will never be presented
with 3D data, k-points or symmetry operations, is simply

\begin{verbatim}
void pdb_read(FILE* infile, struct unit_cell *c,
              struct contents *m);
\end{verbatim}

The current emphasis on {\sc Castep} reflects the research
interests of the author's group, and not a fundamental permanent
limitation of c2x.

For portability the code depends on as few libraries as possible, and
therefore it includes its own FFT routine, based on the GPFA
algorithm\cite{GPFA} with support for any prime factorisation (rather
than being limited to factors of 2, 3 and 5 as {\sc Castep}'s version
of this algorithm is). This FFT code is not optimised, but run time is
not expected to be significant. It would be easy to replace this FFT
with an optimised library if it became necessary.

The one optional library which is supported and recommended is
spglib\cite{spglib}. Without this c2x loses its ability to
perform symmetry analysis. Spglib is also written in C.

As there are so few dependencies, and the code is in standard C, the
build system is simply a makefile. There is a growing test suite to
help ensure the correctness of the code, and to ensure that its
continued development does not introduce regressions.

\section{Conclusion}

C2x is intended to add to the existing ecosystem of utilities used by
the electronic structure community, an ecosystem which includes
OpenBabel and python's ASE. OpenBabel simply converts between a large
number of different formats of cell descriptions. C2x supports fewer
formats, but also includes conversion of density datasets, along with
analysis including integration, interpolation, and line and point
extraction. It also performs conversions to supercells (or the
reverse), and, when linked with spglib, provides symmetry
analysis. C2x is currently simply a command-line utility, so is
callable from most scripting languages. It is used in the AIRSS
project as a command-line interface to spglib.

C2x has a particular emphasis on {\sc Castep}, and is, as far as the
author is aware, the only open-source project which can read {\sc
  Castep}'s binary {\tt .check} files, from which it can extract
atomic positions, charge and spin densities, k-points, symmetry
operations, and wavefunctions. It is intended that this {\tt .check}
file reader could be included in other GPLed code. However, c2x can
also read input and output files from other electronic structure
codes, notably specific formats from {\sc Vasp} and {\sc Onetep},
along with the generic Gaussian cube format.

C2x has particular utility in permitting visualisation and analysis
tools written for one electronic structure code to be used with
another, as it is able to convert charge and spin densities between
different formats. It remains under development, and further
functionality, including closer support for other codes, is likely to
be added.

\section{Acknowledgements}

Over the many years that c2x has been in use, many people have
contributed ideas to it, and it is not possible to list them
all. Particular mention should be made of the TCM Group in the
Cavendish Laboratory at Cambridge for providing an ideal environment
for this project. This work was supported by EPSRC
grant numbers EP/J017639/1, EP/P003532/1 and EP/M011925/1.

\appendix
\section{Example of Defect Production}

\begin{figure}
\begin{center}
\includegraphics[width=0.45\textwidth]{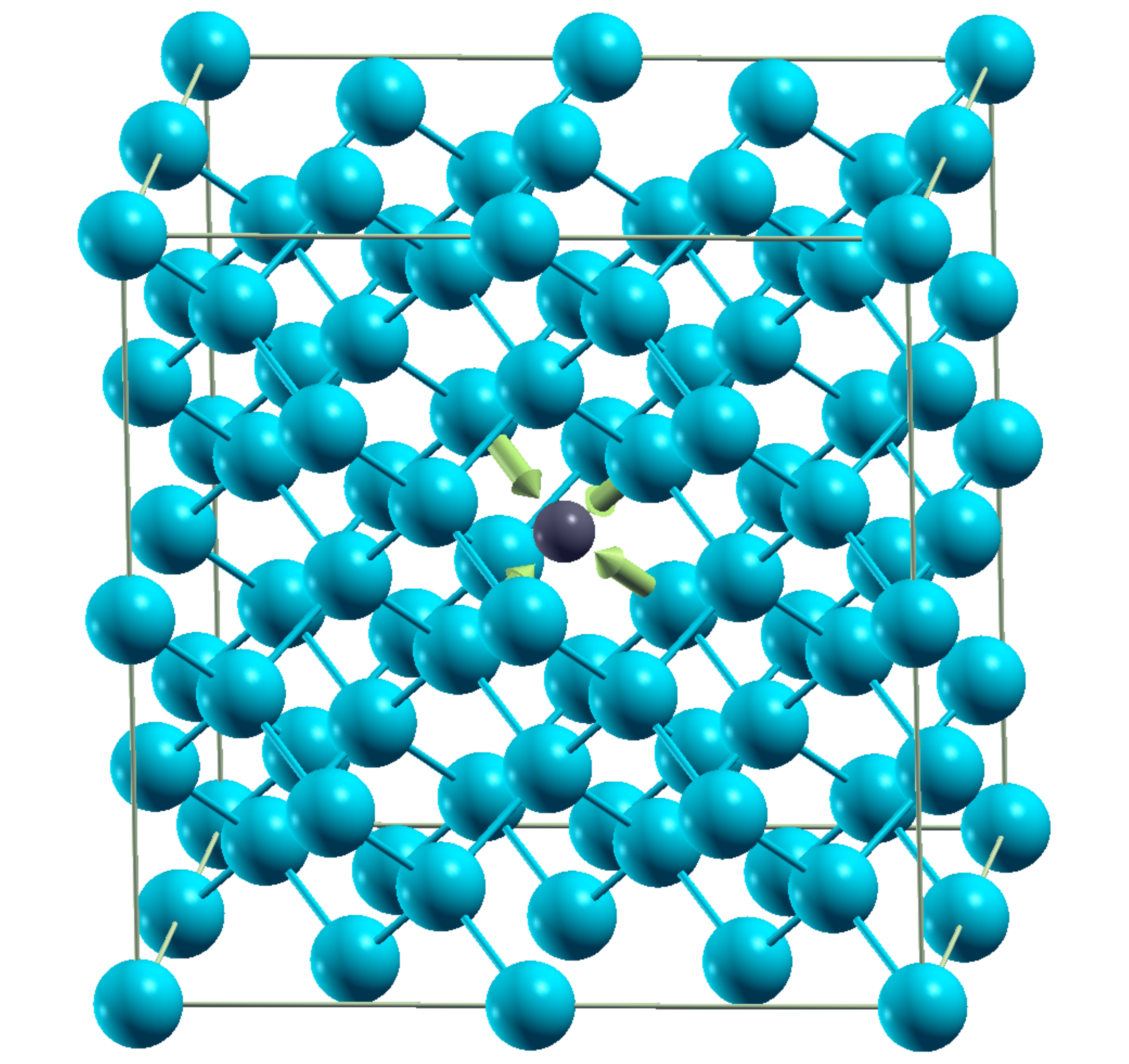}
\end{center}
\caption{Forces, shown as green arrows on the central four cyan Si
  atoms, resulting from a C defect in 64-atom Si cell.
\label{fig:Si63C}
}
\end{figure}

To show the utility of c2x in manipulating input files, the
following example is given. The intention is to show the forces around
a carbon defect in an unrelaxed silicon lattice, starting from a
primitive cell given in {\tt .cell} form as

\begin{verbatim}
%block LATTICE_CART
2.73  2.73 0.00
2.73  0.00 2.73
0.00  2.73 2.73
%endblock  LATTICE_CART

%block POSITIONS_FRAC
Si 0.0     0.0     0.0
Si 0.25    0.25    0.25
%endblock POSITIONS_FRAC
\end{verbatim}

This two-atom cell could be transformed to a cubic eight-atom cell
with the command

\begin{verbatim}
c2x -x='(1,1,-1)(1,-1,1)(-1,1,1)' --cell si2.cell si8.cell
\end{verbatim}

where the {\tt -x} option gives the new cell axes in terms of the
old. Here a 64 atom cubic cell is desired, so the ones in the above
line are replaced by twos. The resulting {\tt .cell} file has its Si
atoms sorted by $c$ co-ordinate, then $b$ co-ordinate, then $a$
co-ordinate, so one can readily find the line

\begin{verbatim}
 Si  0.500000000  0.500000000  0.500000000
\end{verbatim}

and replace `Si' with `C' to create a single central carbon
defect. This file is sufficient input for {\sc Castep} which will
default to calculating the electronic structure with no atomic relaxation.

Running c2x with no options on a {\tt .check} file will produce
an XSF file including forces, and then XCrysden can be used to
visualise the result. The result is shown in figure~\ref{fig:Si63C}
which shows the four silicon atoms surrounding the smaller carbon atom
are pulled towards the carbon.

The symmetry of this unrelaxed defect cell can be obtained with

\begin{verbatim}
$ c2x --int --list Si63C.cell
Tol=0.0001 International symmetry is P-43m
Identity
-4  axis along ( 0.000, 0.000, 1.000) through (0,0,0)
 2  axis along ( 0.000, 0.000, 1.000) through (0,0,0)
-4  axis along ( 0.000, 0.000, 1.000) through (0,0,0)
 2  axis along ( 1.000, 0.000, 0.000) through (0,0,0)
-2  axis along ( 1.000,-1.000, 0.000) through (0,0,0)
 2  axis along ( 0.000, 1.000, 0.000) through (0,0,0)
[output truncated]
\end{verbatim}

Note that a single $\bar{4}$ axis will result in three symmetry
matrices corresponding to applying the rotation once, twice and
thrice, and that the double application of a $\bar{4}$ axis is a $2$ axis.





\bibliographystyle{elsarticle-num}
\bibliography{c2x}







\end{document}